\documentclass[useAMS,usenatbib,letterpaper]{mn2e}

\usepackage{graphicx,amssymb}
\usepackage[top=0.75in,bottom=0.75in,left=0.75in,right=0.75in]{geometry}
\usepackage{verbatim}

\title[Accretion disc wind variability in GRS 1915+105]{Accretion Disc
  Wind Variability in the States of the Microquasar GRS 1915+105}
\author[Neilsen et al.]{Joseph Neilsen,$^{1,2,3}$\thanks{E-mail:
jneilsen@space.mit.edu} Andrew J.\ Petschek,$^{2}$ and Julia
  C.\ Lee$^{2,3}$\\ 
$^{1}$MIT Kavli Institute for Astrophysics and Space Research,
  Cambridge, MA 02139\\
$^{2}$Astronomy Department, Harvard University, Cambridge,
  MA 02138\\
$^{3}$Harvard-Smithsonian Centre for Astrophysics, Cambridge, MA 02138}
\begin{document}

\date{Accepted 2011 December 5. Received 2011 November 22; in original
  form 2011 October 2}

\pagerange{\pageref{firstpage}--\pageref{lastpage}} \pubyear{2011}

\maketitle

\label{firstpage}

\begin{abstract}
Continuing our study of the role and evolution of accretion disc winds in the microquasar GRS 1915+105, we present high-resolution spectral variability analysis of the $\beta$ and $\gamma$ states with the \textit{Chandra} High Energy Transmission Grating Spectrometer. By tracking changes in the absorption lines from the accretion disc wind, we find new evidence that radiation links the inner and outer accretion discs on a range of time-scales. As the central X-ray flux rises during the high-luminosity $\gamma$ state, we observe the progressive over-ionization of the wind. In the $\beta$ state, we argue that changes in the inner disc leading to the ejection of a transient `baby jet' also quench the highly-ionized wind from the outer disc. Our analysis reveals how the state, structure, and X-ray luminosity of the inner accretion disc all conspire to drive the formation and variability of highly-ionized accretion disc winds.
\end{abstract}


\section{Introduction}
\label{sec:intro}
In the last two decades, \textit{ASCA, Chandra,~XMM-Newton,} and
\textit{Suzaku} have discovered a multitude of highly ionized
absorbers in the X-ray spectra of black hole and neutron star X-ray
binaries. In some cases, without clear evidence for blueshifts
(e.g.\ \citealt{Ebisawa97a,Kotani97a,K00,Kotani2000a,L02,M04a,Xiang09}), these 
absorbers have been attributed to hot or extended disc atmospheres. Many 
observations, however, provide unambiguous evidence for outflows from
the accretion disc (e.g.\ \citealt{BS00,Schulz02,U04,M06a,M06b,M08};
\citealt*{U09}; \citealt{NL09,Blum10,ReynoldsM10,Miller11};
\citealt*{N11a}, hereafter Paper II). These results suggest that
accretion disc winds may be as common as relativistic jets in
accreting systems. 
\defcitealias{NL09}{Paper I}
\defcitealias{N11a}{Paper II}
\defcitealias{B00}{B00}

Of particular interest, then, is the variability of these disc winds
and their role in the evolution of the accretion flow. Ionized
absorbers in X-ray binaries are now known to vary on time-scales from 5
seconds (GRS 1915+105, \citetalias{N11a}) to 300 s (H 1743-322,
\citealt{M06b}) to ks or more (Cir X-1, \citealt{Schulz02}; GRS
1915+105, \citealt{L02,U10}). Furthermore, there is mounting evidence
that this variability cannot be due solely to changes in the ionizing
luminosity (\citealt{L02,Schulz02,Blum10}; \citetalias{N11a}), which
implies that high-spectral-resolution X-ray observations of accreting
X-ray binaries may be able to probe either the physical processes that
link outflows to the behavior of the inner accretion flow, or
inhomogeneities in the structure and density of accretion disc winds.

In a study of 10 years of \textit{Chandra} High-Energy Transmission
Grating Spectrometer (HETGS; \citealt{C05}) observations of GRS
1915+105, we demonstrated that the strength of absorption lines in the
accretion disc wind is anticorrelated with the fractional hard X-ray
flux (and the strength of the jet; \citealt{NL09}, hereafter
\citetalias{NL09}; for a preliminary discussion of the absorption
lines, see \citealt{M08}). Based on this anticorrelation and estimates
of the mass loss rates in the wind and the jet, we argued that
accretion disc winds may be able to suppress or quench jets by
directly draining their matter supply on long time-scales. The
fractional hard X-ray flux is therefore a useful indicator of both the
accretion state and the physics of outflows around black holes.
\begin{figure*}
\centerline{\includegraphics[width=0.75\textwidth]{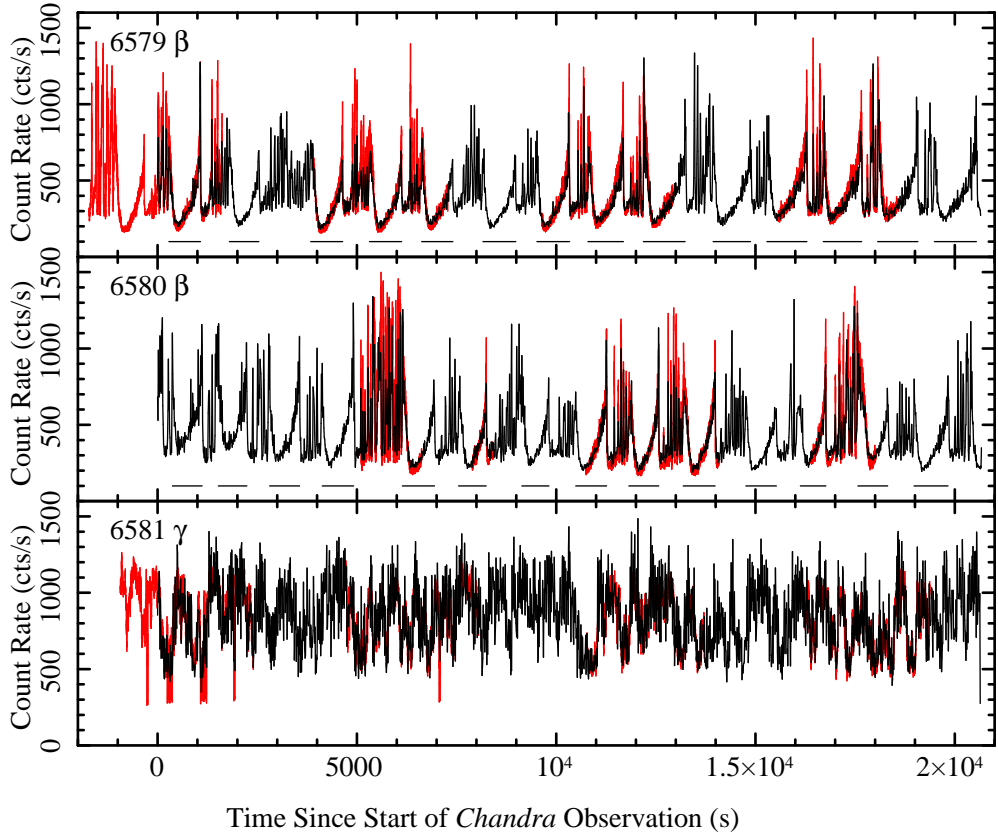}}
\caption{The \textit{Chandra} (black) and \textit{RXTE} (red)
  lightcurves of the $\beta$ (top, middle) and $\gamma$ (bottom)
  states. The \textit{RXTE} lightcurves have been renormalized to
  match the mean \textit{HETGS} count rate, with excellent
  agreement between the two missions. In the top and middle panels, we
  use horizontal lines to indicate the locations of the X-ray dips
  analysed in Section \ref{sec:betahetgs}.} 
\label{fig:lc}
\end{figure*}
Yet much more can be said for GRS 1915+105, whose X-ray lightcurve
exhibits at least 14 different patterns of variability (\citealt{B00},
hereafter B00; \citealt{K02,Hannikainen05}). Many of these
phenomenological patterns, which are highly structured and
high-amplitude, are interpreted as limit cycles of accretion and
ejection in an unstable disc \citep{B97b,M98,T04,FB04}. In order to
understand the physics linking disc winds, jets, and this fast X-ray 
variability, we presented the first detailed phase-resolved spectral
analysis of a variability class in GRS 1915+105, using a joint
\textit{RXTE/Chandra} HETGS observation of the $\rho$ state
\citepalias{N11a}. We showed for the first
time that changes in the broadband X-ray spectrum produce changes in
the \textit{structure} and mass loss rate of the accretion disc wind
on time-scales as short as 5 seconds. Furthermore, our analysis
indicated that the wind in the $\rho$ state may be sufficiently
massive to excite long-term oscillations in the mass of the disc 
(\citealt{Shields86,Luketic10}). By establishing clear causal links
between the X-ray oscillations and the disc wind, \citetalias{N11a} 
provides a physical mechanism that could be responsible for both the
suppression of jets and (possibly) state transitions in black hole
binaries. 

In this paper, we continue our follow-up wind variability studies with
spectral timing analysis of three additional archival \textit{Chandra}
HETGS observations of GRS 1915+105. Two of these observations (ObsIDs
6579 and 6580) fall into the $\beta$ class, a wild $\sim30$-min cycle
that is associated with the production of impulsive/discrete
(``baby'') radio jets
(\citealt{PF97,Fender97,M98,E98a,Mirabel99,B00,K02,FB04}). Each
$\beta$ cycle begins with a 10--15 minute dip in the lightcurve, which 
is  spectrally hard and exhibits a QPO whose frequency tracks the
X-ray flux (e.g.\ \citealt{Markwardt99,Mikles06,Rodriguez08b}). The
dip ends with a bright spike, followed by 10--15 minutes of soft,
bright X-ray flares. These flares are coincident with strong
optically-thin radio/infrared oscillations (baby jets), but there is
an ongoing debate as to whether these discrete ejections are produced
during the dip itself or at the time of the X-ray spike
\citep{M98,E98a,Migliari03,Rothstein05,Rodriguez08a}. 
\setcounter{table}{0}
\begin{table}
\caption{\textit{Chandra} HETGS observations of GRS 1915+105}
\label{tbl:obs}
\begin{tabular}{@{}ccccc}
\hline
Obs.       & X-ray  &               & Elapsed    & T$_{\rm exp}$ \\
ID         & State  & Date          & Time (ks)  & (ks)\\
\hline
6579       & $\beta$ & 2005 Dec.\ 1 & 20.65 & 12.30 \\
6580       & $\beta$ & 2005 Dec.\ 1 & 21.62 & 12.14 \\
6581       & $\gamma$ & 2005 Dec.\ 3 & 20.65 & ~9.73 \\
\hline
\end{tabular}
\end{table}

During the third observation (ObsID 6581), GRS 1915+105 exhibited
irregular variability with very high X-ray flux (\textit{RXTE} count 
rate $\ga6000$ counts s$^{-1}$ PCU$^{-1}$) and extremely strong
red noise in the power density spectrum ($P_{\nu}\sim\nu^{-2}$). 
Although we identified this as the $\gamma$ state in \citetalias{NL09}
and continue to refer to it thus here, this particular observation is
rather unusual since it also displays significant similarities to
typical $\delta$ and $\mu$ states. Relatively little has been said
about the $\mu,~\delta,$ and $\gamma$ states in the literature, as
none have been associated with strong radio jets \citep{K02}. However,
in a study of the $\mu$ state, \citet*{Soleri08} reported the
discovery of a transient low-frequency QPO, which they argue may be
linked to state transitions or transient jets. This leaves open the
possibility that the $\mu/\delta/\gamma$ states exhibit
physically-interesting outflow behavior, despite their weak radio
flux. In this respect, these states could be similar to the $\rho$
state, which has very low radio brightness \citep{K02} but exhibits
extreme disc wind variability and may produce short-lived jets 
\citepalias{N11a}.

Here we address the variability of the known accretion disc wind in
the \textit{Chandra} HETGS observations of the $\beta$ and 
$\gamma$ states, whose lightcurves are shown in Figure
\ref{fig:lc}. By comparing changes in the ionized absorption to
changes in the X-ray flux, we show how the wind is linked to the
oscillations in the inner disc. The paper is organized as follows: in
Section \ref{sec:obs} we describe our observations and data
reduction. In Section \ref{sec:beta} we perform spectral analysis of
the wind in the $\beta$ state with the \textit{Chandra} HETGS and
\textit{RXTE} PCA; in \ref{sec:gamma} we use the HETGS to study the
wind in the $\gamma$ state. We discuss the implications of our
results in Section \ref{sec:discuss}.\vspace{-7mm}

\section{OSERVATIONS AND DATA REDUCTION}
\label{sec:obs}
GRS 1915+105 was observed with the \textit{Chandra}~HETGS on 2005
December 1 (01:45:09 UT and 18:41:30 UT) and Dec 3 (16:26:06 UT) for
20.65 ks, 21.62 ks, and 20.65 ks, respectively. The data were taken in
Continuous Clocking Mode, which has a time resolution of 2.85 ms, in
order to mitigate photon pileup. However, due to the high X-ray flux
the observations suffered from severe telemetry saturation, so that
the good exposure times for these observations were only 12.30 ks,
12.14 ks, and 9.73 ks (details in Table \ref{tbl:obs}). 

We reduce and barycentre-correct the \textit{Chandra}~data using
standard tools from the {\sc ciao} analysis suite, version 4.3. We use
the order-sorting routine to remove the ACIS S4 readout streak, since
the \textit{destreak} tool can introduce spectral artefacts for bright  
continuum sources like GRS 1915+105. After reprocessing and filtering,
we extract High-Energy Grating (HEG) spectra and create grating
responses. For the $\gamma$ observation (ObsID 6581), we find that the
detected position of the $0^{\rm th}$-order image results in an
offset of $-44$ eV between the HEG $-1$ and HEG $+1$ orders (measured
from the centroid of the Fe\,{\sc xxvi} line at $\sim7$ keV). Shifting
the position of the $0^{\rm th}$-order image to the nominal source
position results in an offset of 36 eV, so we take the average 
of the detected position and the nominal position as the true location
of the $0^{\rm th}$-order image on the CCD. The resulting position
error from this average is less than 0.1 pix, so the associated
uncertainties in the wavelength calibration are smaller than our
uncertainty in the line centroids. We also extract 10-second
lightcurves with \textit{dmextract}.  

Due to incomplete calibration of Charge-Transfer Inefficiency (CTI) in
CC mode, there is some wavelength-dependent disagreement in the
continuum flux between spectral orders of the HEG (and the MEG, which
we do not consider here because of its lower spectral resolution). For
this reason, it is not currently possible to fit a physical continuum
model to the HETGS data. Instead, we fit the individual spectra with
polynomials to model the local continuum and use Gaussians for line
features found in the combined residuals. 

During the \textit{Chandra}~observations, \textit{RXTE} made pointed
observations of GRS 1915+105, lasting 23.1 ks, 13.7 ks, and 21.7 ks
(elapsed), with exposure times of 13.2 ks, 7.2 ks, and 13.6 ks,
respectively. We select all available data subject to the following
constraints: (1) the Earth-limb elevation angle is above 3$^\circ$;
(2) the spacecraft is outside the South Atlantic Anomaly; (3) the
offset angle from GRS 1915+105 is less than 0.02$^\circ$. In this
paper, we analyse the data from the Proportional Counter Array (PCA),
which covers the 2--60 keV band, in order to verify and supplement the 
variability detected by \textit{Chandra}. 
For timing analysis, we make use of the data from the binned mode
B\_8ms\_16A\_0\_35\_H\_4P, which covers the 2.0--14.8 keV band at 7.8 
ms time resolution, and the event mode E\_16us\_16B\_36\_1s, which
covers the 14.8--60 keV band at 15.3 $\mu s$ time resolution. We
extract and combine 1-second background subtracted lightcurves from
each of these modes. For comparison to the \textit{Chandra} data, we
barycentre the \textit{RXTE} lightcurves. For spectral analysis, we
extract Standard-2 129-channel spectra from relevant time intervals
(see Section \ref{sec:beta} and Figure \ref{fig:lc}). We restrict our 
analysis to the 3--45 keV top layer spectrum of PCU-2, which is best
calibrated. We assume 0.6\% systematic errors. Although these
states exhibit some strong variability on time-scales shorter than the
16-second bins of the Standard-2 data, our goal is to use the spectra
to understand the ionization of the wind, not to characterize physical
variability in the accretion disc and corona. \vspace{-5mm} 

\section{The $\beta$ State}
\label{sec:beta}
\begin{figure}
\centerline{\includegraphics[width=3.0 in]{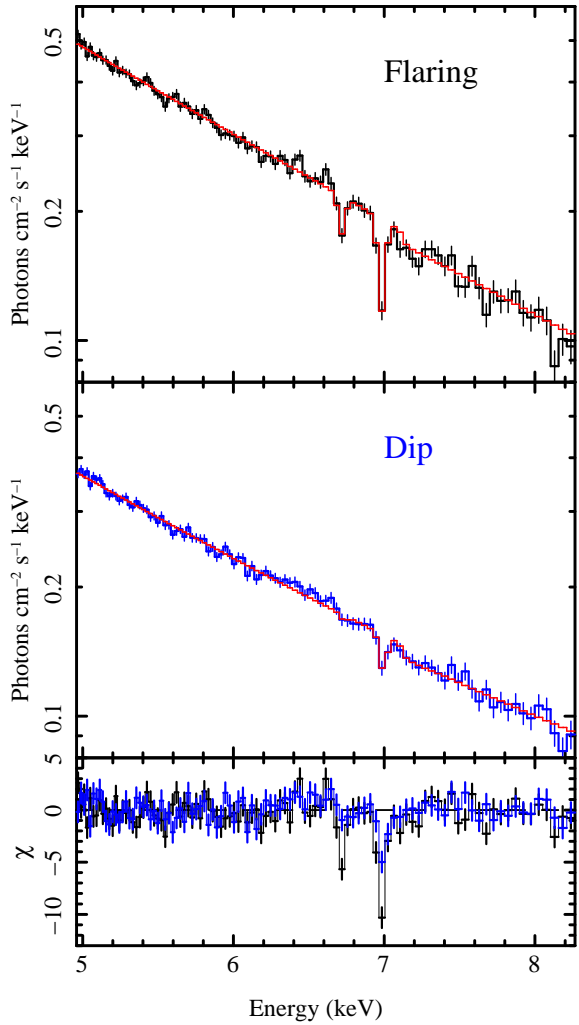}}
\caption{The \textit{Chandra} HETGS spectra and residuals for the
  flaring (top panel) and dip (middle panel) intervals of the $\beta$
  state of GRS 1915+105. We detect strong absorption lines from
  Fe\,{\sc xxvi} during both time periods, and Fe\,{\sc xxv} during 
  the X-ray flaring. These spectra include portions of
  observations 6579 and 6580, coadded to improve our sensitivity. The
  residuals are calculated with the line normalizations set to
  zero.\vspace{-3mm}}
\label{fig:betaspec}
\end{figure}
In this section, we explore the variability and dynamical evolution of
the accretion disc wind in the $\beta$ state. Here, links between the
extreme X-ray variability and the disc wind are particularly
interesting given the association of the $\beta$ state with
transient jets and disc-jet interactions. Previously, we
analysed the average spectrum of these two (ObsIDs 6579 and 6580;
Table \ref{tbl:obs}) joint RXTE/Chandra observations of GRS 1915+105
in the $\beta$ state (Paper I). We found the average RXTE continuum
during each of these observations to be relatively soft, with
$\sim79\%$ of the 3--18 keV X-ray luminosity emitted below 8.6 keV  
($L_{\rm X}\sim6\times10^{38}$ ergs s$^{-1}$). In both time-averaged
high-resolution X-ray spectra from the HETGS, we detected an 
Fe\,{\sc xxvi} Ly$\alpha$ absorption line 
from the accretion disc wind. This feature had equivalent widths of
$W_{0}=13.3^{+3.0}_{-2.9}$ eV and $19.3^{+3.2}_{-3.5}$ eV in observations
6579 and 6580, respectively; the corresponding blueshifts were
$910^{+430}_{-390}$ km s$^{-1}$ and $1100^{+300}_{-360}$ km s$^{-1}$.
\begin{figure}
\centerline{\includegraphics[width=3.2 in]{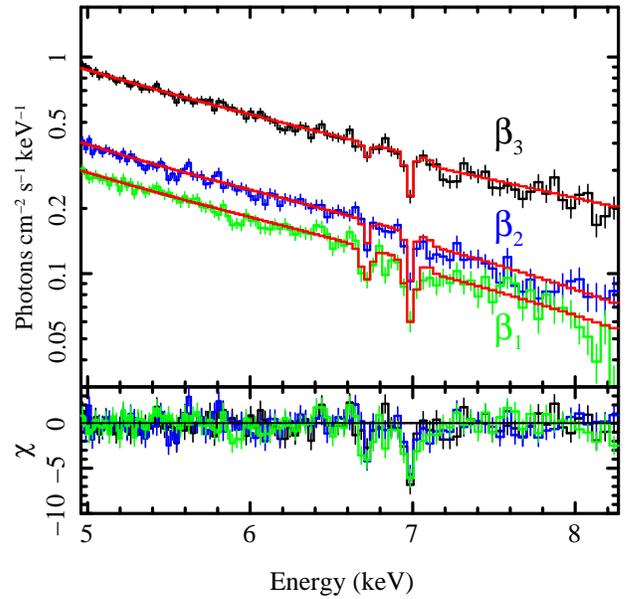}}
\caption{\textit{Chandra} HETGS flux-resolved spectra and
  residuals for the flaring interval of the $\beta$ state. As
  discussed in the text, the weakening of the Fe\,{\sc xxv} line can
  be taken as an indication that the ionization of the wind may increase
  slightly with the continuum flux during the flares.\vspace{-6mm}}
\label{fig:betaburst}
\end{figure}
\begin{table*}
 \centering
 \begin{minipage}{157mm}
  \caption{X-ray absorption lines in GRS 1915+105\label{table:abs}}
  \begin{tabular}{@{}ccclcccccc@{}}
  \hline
  &
  &
  &
  &
$E_{0}$  &
$E_{\rm obs}$  &
$\Delta v_{\rm shift}$  & 
$\sigma_{\rm v}$  & 
  & 
$W_{0}$ \\
State  &
Interval&
$F_{5-8}$  &
Line  &
(keV)  &
(keV)  &
(km s$^{-1}$)  &
(km s$^{-1}$)  &
Flux  &
(eV)\\
 \hline
 \vspace{1mm}$\beta$ & Dip &$6.0\pm0.5$& Fe\,{\sc xxvi} & 6.961 & $6.993^{+0.009}_{-0.012}$ & $-1340_{-400}^{+520}$ 
& $<1310$& $1.9^{+0.7}_{-0.6}$& $11.3_{-3.8}^{+4.1}$\\
\vspace{1mm} &  && Fe\,{\sc xxv} & 6.700 & $6.730$\footnote{Since the Fe\,{\sc xxv} line is not very significant
  during this interval, we tie its velocity to that of Fe\,{\sc xxvi}.} & $-1340^{a}$
&  = Fe\,{\sc xxvi}& $<0.8$ & $<4.8$\\
\vspace{1mm} & Flare &$7.6\pm0.4$& Fe\,{\sc xxvi} & 6.961 & $6.982^{+0.006}_{-0.005}$ &$-900_{-240}^{+230}$ & $660^{+230}_{-240}$
& $5.4\pm0.8$& $26.0_{-3.9}^{+3.8}$\\
\vspace{1mm} &  && Fe\,{\sc xxv} & 6.700 & $6.718^{+0.010}_{-0.008}$ &  $-780_{-450}^{+360}$ &  = Fe\,{\sc xxvi}
 & $2.7\pm0.7$& $11.4_{-3.0}^{+2.8}$\\
\vspace{1mm}&$\beta_{1}$&$4.6\pm0.4$&Fe\,{\sc xxvi}&6.961&$6.986^{+0.009}_{-0.010}$&$-1060_{-390}^{+420}$&$1140^{+360}_{-300}$
& $5.2\pm0.1$& $42.2_{-8.1}^{+8.1}$\\
\vspace{1mm}&&&Fe\,{\sc xxv}&6.700&$6.71\pm0.01$&$-560_{-590}^{+470}$&=
Fe\,{\sc xxvi} & $3.5^{+0.9}_{-1.0}$& $24.6_{-6.9}^{+6.6}$\\
\vspace{1mm}&$\beta_{2}$&$6.1\pm0.5$&Fe\,{\sc xxvi}&6.961&$6.984^{+0.012}_{-0.007}$&$-950_{-520}^{+310}$&$<790$ & $3.8\pm0.1$& $22.9_{-6.2}^{+5.9}$\\
\vspace{1mm}&&&Fe\,{\sc xxv}&6.700&$6.718^{+0.013}_{-0.009}$&$-800_{-560}^{+390}$& =
Fe\,{\sc xxvi} & $2.1\pm0.9$& $11.3_{-4.8}^{+4.8}$\\
\vspace{1mm}&$\beta_{3}$&$13.9\pm0.3$&Fe\,{\sc xxvi}&6.961&$6.980^{+0.008}_{-0.005}$&$-780_{-360}^{+220}$&$<910$ & $8.6^{+1.9}_{-2.0}$& $22.3_{-5.1}^{+4.9}$\\
\vspace{1mm}&&&Fe\,{\sc xxv}&6.700&$6.718^{a}$&$-780^{a}$& =
Fe\,{\sc xxvi} & $3.2^{+1.7}_{-1.8}$& $7.5_{-4.1}^{+3.9}$\\

\vspace{1mm}$\gamma$&$\gamma_{1}$&$18.8\pm1.5$& Fe\,{\sc xxvi} & 6.961 & $6.985\pm0.008$ & $-1020_{-330}^{+320}$ 
& $700^{+410}_{-430}$& $11.6\pm2.8$& $21.3\pm5.1$\\
\vspace{1mm}&&& Fe\,{\sc xxv} & 6.700 & $6.723^{a}$ & = Fe\,{\sc xxvi} 
& = Fe\,{\sc xxvi}& $3.5^{+2.2}_{-2.3}$& $5.9_{-3.9}^{+3.7}$\\
\vspace{1mm}&$\gamma_{2}$&$22.5\pm2.0$& Fe\,{\sc xxvi} & 6.961 & $6.983^{+0.009}_{-0.008}$ & $-920_{-390}^{+350}$ 
& $980^{+540}_{-410}$& $14.1^{+3.4}_{-3.0}$& $21.4_{-4.6}^{+5.2}$\\
\vspace{1mm}&&& Fe\,{\sc xxv} & 6.700 & $6.721^{a}$ & = Fe\,{\sc xxvi}
& = Fe\,{\sc xxvi}& $<5.7$& $<7.9$\\
\vspace{1mm}&$\gamma_{3}$&$24.8\pm2.2$& Fe\,{\sc xxvi} & 6.961 & $6.990\pm0.007$ & $-1240\pm300$ 
& $1000^{+420}_{-350}$& $19.0^{+3.7}_{-3.4}$& $26.4_{-4.8}^{+5.1}$\\
\vspace{1mm}&&& Fe\,{\sc xxv} & 6.700 & $6.728^{a}$ & = Fe\,{\sc xxvi}
& = Fe\,{\sc xxvi}& $<4.5$& $<5.7$\\
\vspace{1mm}&$\gamma_{4}$&$26.4\pm2.3$& Fe\,{\sc xxvi} & 6.961 & $6.998\pm0.007$ & $-1550_{-280}^{+300}$ 
& $790^{+490}_{-680}$& $17.6^{+3.9}_{-3.7}$& $23.1_{-4.8}^{+5.1}$\\
\vspace{1mm}&&& Fe\,{\sc xxv} & 6.700 & $6.735^{a}$ & = Fe\,{\sc xxvi} 
& = Fe\,{\sc xxvi}& $<6.6$& $<7.6$\\
\vspace{1mm}&$\gamma_{5}$&$28.3\pm2.2$& Fe\,{\sc xxvi} & 6.961 & $6.996^{+0.008}_{-0.010}$ & $-1490_{-340}^{+440}$ & $<1330$& $12.9^{+2.8}_{-3.2}$& $16.0_{-4.0}^{+3.5}$\\
\vspace{1mm}&&& Fe\,{\sc xxv} & 6.700 & $6.734^{a}$ &  = Fe\,{\sc xxvi} 
&  = Fe\,{\sc xxvi} & $<6.5$& $<7.4$\\
\hline
\end{tabular}
Errors quoted are 90\% confidence ranges for a single
  parameter unless otherwise stated. State: X-ray variability type in
  the classification of \citetalias{B00}. Interval: line parameters
  are reported for different flux levels and time intervals (see text
  for more details). $F_{5-8}:$ 5--8 keV (absorbed) continuum flux in
  units of $10^{-9}$ photons s$^{-1}$ cm$^{-2}$, with errors given by
  the standard deviation between the HEG $\pm$1 orders; $E_{0}:$ rest
  energy; $E_{\rm obs}:$ measured energy; $\Delta v_{\rm shift}:$
  measured Doppler velocity; $\sigma_{\rm v}$: line width; Flux:
  measured absorbed line flux in units of 10$^{-3}$ photons s$^{-1}$
  cm$^{-2};~W_{0}:$ line equivalent width.\vspace{-5mm}\end{minipage}
\end{table*}
In what follows, we use the known properties of the $\beta$ state to
extract high-resolution X-ray spectra corresponding to
physically-interesting intervals of this unusual oscillation. We
compensate for severe telemetry saturation (leading to reduced counts,
Section \ref{sec:obs}; Table \ref{tbl:obs}) by fitting both
observations 6579 and 6580 jointly. All spectral fitting is done from
5--8 keV in ISIS \citep{HD00,Houck02}. We assume a distance and
inclination of $D=11.2$ kpc and $i=66^{\circ}$ \citep{F99}; we fix
$N_{\rm H}=5\times 10^{22}$ cm$^{-2}$ (\citealt{L02} and references
therein).

In general, choosing time intervals for spectral timing analysis is
difficult, but the $\beta$ lightcurve is easily divided into two
sections with different physical properties: (1) the long, hard,
discrete jet-producing dip and (2) the subsequent period of
bright flaring. Examination of the $\beta$-state X-ray lightcurves in
the literature and in our data indicates that the hard dip is
bracketed on the left by the last bright flare of the preceding
flaring interval, and on the right by peak or falling edge of the
X-ray spike. Using the \textit{Chandra} lightcurve and the PCA
lightcurve where available, we hand-select the dip and flare intervals
accordingly. The resulting GTIs are shown underneath the X-ray
lightcurves in the top and middle panels of Figure \ref{fig:lc}.\vspace{-3mm}

\subsection{\textit{Chandra} HETGS}
\label{sec:betahetgs}
Applying these GTIs to the \textit{Chandra} observations, we extract
high-resolution spectra corresponding to the dip and flaring
intervals. We show the resulting spectra in Figure
\ref{fig:betaspec}. As discussed in Section \ref{sec:obs}, we model the
\textit{Chandra} X-ray continuum with polynomials to account for
calibration uncertainties. This technique effectively isolates narrow
absorption lines by characterizing the local continuum. We use
Gaussian lines to measure the properties of the ionized iron
absorption lines during each time interval; our results are shown
in Table \ref{table:abs}.

Both Figure \ref{fig:betaspec} and Table \ref{table:abs} reveal clear
differences between the accretion disc wind during the flare and dip
intervals. From Figure \ref{fig:betaspec}, it can be seen that during
the flares (top panel), we detect strong absorption lines from
both Fe\,{\sc xxvi} (Ly$\alpha$: 1s -- 2p, 6.96 keV) and Fe\,{\sc xxv}  
(He$\alpha$: 1s$^{2}$ -- 1s2p, 6.7 keV). The significance of the
Fe\,{\sc xxv} and Fe\,{\sc xxvi} lines ($\sim6\sigma$ and
$\sim11\sigma$) indicates a relatively strong, highly-ionized
outflow. During the dip (middle panel), Fe\,{\sc xxvi} is much weaker
and Fe\,{\sc xxv} is not significant at all, although it can be seen
as a faint blip in the residuals (bottom panel). Although the 5--8 keV
continuum flux decreases (Table \ref{table:abs}) from the flares to
the dip, the bolometric luminosity remains roughly constant (Section
\ref{sec:betapca}). 

Interestingly, Table \ref{table:abs} shows that aside from the
reduction in line flux, there are few if any quantitative differences 
between the iron absorbers during these two parts of the $\beta$
cycle. During the flares, the wind is blueshifted by $900^{+240}_{-230}$ km
s$^{-1}$ and has a line-of-sight velocity dispersion of $\sigma_{\rm v} 
=660_{-240}^{+230}$ km s$^{-1}.$ In the spectrally-hard dip, the
blueshift is $1340^{+400}_{-520}$ km s$^{-1}$ and the lines are
unresolved, but we constrain the velocity dispersion to be $<1310$ km
s$^{-1}.$ These results are consistent with the long-term behavior of
the wind that we reported in \citetalias{NL09}. Perhaps the most
surprising result, however, is that despite the low significance of
Fe\,{\sc xxv} in the dip, the dip and flare spectra are statistically 
consistent with a constant Fe\,{\sc xxv}/Fe\,{\sc xxvi} ratio. This
suggests that the ionization state of the plasma in the wind might not
vary from the dip to the flares, despite known changes in the hardness
of the X-ray spectrum (Section \ref{sec:betapca}).

We can substantiate this claim using a simple multiplicative Gaussian
model based on chapter 9 of \citet{Draine11} that replaces the line
flux parameter with the ion column density. First, we fit both phases
of the cycle separately, allowing the Fe\,{\sc xxvi}/Fe\,{\sc xxv}
ratio to vary between the dip and the flares. Then we fit them
together, requiring both intervals to have the same 
Fe\,{\sc xxvi}/Fe\,{\sc xxv} ratio and line width. With the line 
ratios free, we achieve $\chi^{2}/\nu=884.0/791=1.12;$ with the line
ratios and widths tied together, we obtain the same reduced
$\chi^{2}=885.8/793 =1.12.$ In other words, there is no evidence (at
the level of our line detections) that the ionization state of the
wind changes between the dip and the flare. 

Using the results from our ``tied'' model, we find that the average
Fe\,{\sc xxvi}/Fe\,{\sc xxv} ratio is $r=0.17_{-0.05}^{+0.06};$ the
Fe\,{\sc xxvi} column density is $N_{\rm Fe\,26}=2.6_{-0.8}^{+0.9} 
\times10^{17}$ cm$^{-2}$ during the dip and $N_{\rm Fe\,26}=
8.0_{-1.3}^{+1.6}\times 10^{17}$ cm$^{-2}$ during the flares. To
convert from the ion column density to equivalent hydrogen columns, we
assume elemental abundances from \citet*{Wilms00}. Assuming only the
ionic abundances of Fe\,{\sc xxv} and Fe\,{\sc xxvi} are
non-zero, 
then these results imply 
an equivalent hydrogen column density in the wind of $N_{\rm H}=1.0
\pm0.3\times10^{22}$ cm$^{-2}$ during the dip and $N_{\rm H}=
3.0_{-0.5}^{+0.6}\times10^{22}$ cm$^{-2}$ during the flares. We
obtained similar results with photoionization models of the disc wind
in the $\rho$ state \citepalias{N11a}, so we believe our simple
measure of $N_{\rm H}$ here is plausible. We conclude that the column
density of the wind decreases significantly from the flares to the
dip, while its bulk dynamical and ionization properties remain
constant.
 
We can also ask whether or not the absorption lines vary within the
flaring phase (our S/N is too low to perform a similar analysis of
line variability in the dip). Using the \textit{Chandra} lightcurve,
we create new GTI files to divide the flare phase into three flux
bands with roughly equal exposure times. The resulting spectra (see
Figure \ref{fig:betaburst}) reveal significant differences in the wind
between the lowest flux levels and the highest flux levels. Visible
differences include apparent decreases in the equivalent widths and
line widths, and a decrease in the strength of the Fe\,{\sc xxv} line
relative to Fe\,{\sc xxvi} with increasing continuum flux. There is no
evidence for variations in the velocity of the wind, but the changes
in the line width and the Fe\,{\sc xxv}/Fe\,{\sc xxvi} ratio are
marginally significant at 90\% confidence (Table \ref{table:abs}). 
Thus it appears that the ionization state or density of the wind may
be somewhat sensitive to the luminosity on short
time-scales.\vspace{-5mm} 

\begin{figure}
\centerline{\includegraphics[width=3.3 in]{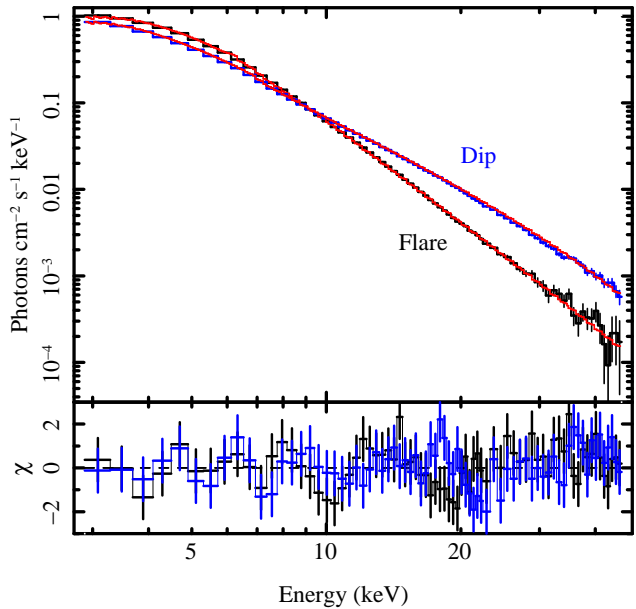}}
\caption{\textit{RXTE} PCA spectra of the $\beta$-state dip and flare,
  for comparison to Figure \ref{fig:betaspec}. The dip spectrum is
  notably harder than the spectrum of the flaring phase, but the
  1 ev -- 1 MeV luminosities are comparable at $10^{39}$ ergs
  s$^{-1}$. This suggests that the changes in the X-ray absorber may
  require variations in the density or geometry of the wind (see
  Section \ref{sec:dipflare}).\vspace{-5mm}} 
\label{fig:betapcaspec}
\end{figure}
\subsection{\textit{RXTE} PCA}
\label{sec:betapca}
With sensitive coverage out to 45 keV, \textit{RXTE} PCA spectra can
also provide valuable insight into variable ionization processes, so
we extract PCA spectra for the dip and flaring phases with the same
GTIs used for the HETGS (see Figure \ref{fig:lc}). Our purpose here is
not to explore the variability of the accretion disc and corona, which
has been done for the $\beta$ state
(e.g.\ \citealt{Migliari03,Rothstein05}), but to capture the average
shape of the ionizing spectrum. With typical wind densities around
$10^{12}$ cm$^{-3}$ \citep{L02,U09,N11a}, the recombination time-scale
is of the order of a few tenths of ms \citep{Kallman09}. Thus our
treatment of the average spectrum is justified because the wind can
effectively respond instantly to changes in the ionizing spectrum. 

We find that the average dip spectrum can be reproduced with a model 
consisting of cold absorption ({\tt tbabs}; \citealt{Wilms00}), a
disc component ({\tt ezdiskbb}; \citealt{Zimmerman05}), a Comptonized
component ({\tt nthcomp}; \citealt*{Zdziarski96,Zycki99}), and an
emission line in the 5--7 keV range ({\tt egauss}). The disc is cool
($\sim0.75$ keV) with a normalization ($\sim800$) that implies an
inner radius around 120 km. The photon index is $\Gamma\sim2.5$, and
the electron temperature is $\sim12$ keV; we tie the seed photon
temperature to the disc temperature. During the flaring interval, we
replace {\tt nthcomp} with a high-energy cutoff ({\tt highecut}) and
{\tt simpl}, a convolution model that takes a seed spectrum and
scatters a fraction of the photons into a power law
\citep{Steiner09a}. Our fits to the average flare spectrum imply a
slightly hotter, smaller disc ($T\sim1.1$ keV, $R\sim100$ km), a
steeper power law ($\Gamma\sim3.8$) comprising $\sim60\%$ of the 
photon flux, and an $e$-folding energy $>75$ keV (we set the cutoff
energy to zero). The fit parameters themselves must be interpreted
cautiously given the strong variability in the flare and dip, but our
models provide a good description of the spectra
($\chi^{2}/\nu=222.9/264$; see Figure \ref{fig:betapcaspec}). 

Again, our primary interest is in the influence of the broadband
continuum on the ionization of the wind, particularly the strength of
the Fe\,{\sc xxv} absorption line. To that end, we integrate our
models over the Fe\,{\sc xxv} photoionization cross-section
\citep{Verner96} to determine the expected ionizing flux. In the
flares, we find $\sim1.3\times10^{-3}$ ionizing photons s$^{-1}$
cm$^{-2}$; during the dip, we find $\sim20\%$ more ionizing 
photons ($\sim1.5\times10^{-3}$ photons s$^{-1}$ cm$^{-2}$). Both
spectra have 0.001--1000 keV luminosities near $10^{39}$ ergs
s$^{-1}$. Since the luminosity and ionizing flux changes are modest,
it is conceivable that the ionization state of the wind remains
constant between the dip and the flares. However, the impact of the
ionizing flux on the lines also depends on the changing density and
geometry of the wind, as well as its characteristic distance from the
continuum source (see \citetalias{N11a} and references therein). We
explore these factors in more detail in Section
\ref{sec:dipflare}. \vspace{-3mm} 

\setcounter{footnote}{0}

\section{The $\gamma$ STATE}
\label{sec:gamma}
In this section, we turn our attention to the variability of the
accretion disc wind in the $\gamma$ state (ObsID 6581; Table
\ref{tbl:obs}). In Paper I, we reported the average \textit{RXTE}
continuum to be very soft, with $\sim82\%$ of the 3--18 keV X-ray
luminosity emitted below 8.6 keV ($L_{\rm X}\sim13\times 10^{38}$ ergs
s$^{-1}$). In the time-averaged \textit{Chandra} HETGS spectrum, we
detected an Fe\,{\sc xxvi} absorption line with an equivalent width of
$-21.9^{+2.2}_{-2.7}$ eV and a blueshift of $1000_{-240}^{+220}$ km
s$^{-1}.$ 

Although no significant jet activity has been reported in the
$\gamma,~\delta$, or $\mu$ states \citep{K02}, the strong red-noise
variability (bottom panel of Figure \ref{fig:lc}) may still
provide clues about the links between radiation and outflows in
accreting X-ray binaries. In what follows, we discuss the variability
of the wind in this state, which is largely unexplored.

The erratic variability here poses some difficulties for grating
studies, since phase-folding (e.g.\ \citealt{N11a}) and filtering on
hardness-hardness-intensity diagrams \citep{Soleri08} are not
possible. These difficulties are compounded by the telemetry
saturation discussed in Section \ref{sec:obs}. The simplest approach,
dividing the observation into several time intervals, reveals no
statistically-significant variability in the wind, so we opt to
extract flux-resolved spectra based on the count rate in the
\textit{Chandra} lightcurve. Because the exposure time and S/N are
already relatively small, we use a sliding box
(e.g.\ \citealt{Schulz02,N11a}) to create five spectra, each covering
roughly one-third of the count rate distribution (with overlap). While
this method alleviates the difficulty in selecting physically-motivated
GTIs, it does create some overlap between consecutive spectra. We only
use statistically-independent spectra for our variability
calculations.
\begin{figure}
\centerline{\includegraphics[angle=270,width=3.1 in]{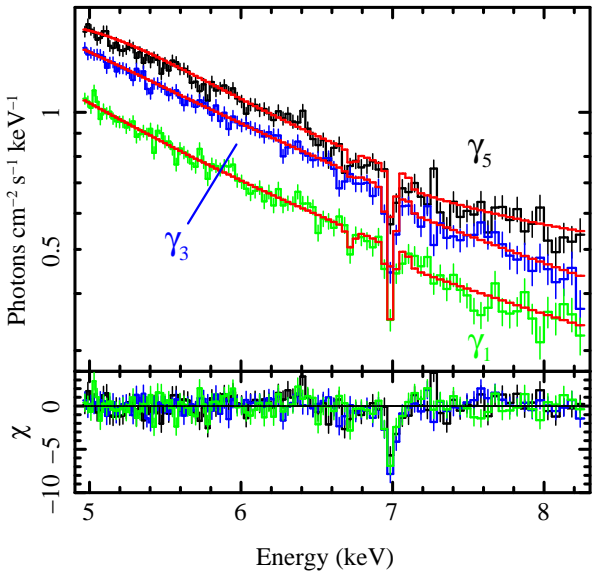}}
\caption{The \textit{Chandra} HETGS flux-resolved spectra and
  residuals for three distinct intervals in the $\gamma$ state. The 
  continuum level changes, but the lines are very similar during these
  intervals. However, the complicated relationship between the
  absorbed line flux and the continuum (Section \ref{sec:gamma})
  indicates that photoionization effects are still important at these
  high luminosities.\vspace{-3mm}}
\label{fig:gammaspec}
\end{figure}

We show the three statistically-independent spectra ($\gamma_{1},
~\gamma_{3},$ and $\gamma_{5}$) in Figure \ref{fig:gammaspec}; all
three spectra contain a strong Fe\,{\sc xxvi} line with a blueshift of 
$1000-1500$ km s$^{-1}.$ In contrast to the $\beta$-state spectra,
here there is little evidence for absorption by Fe\,{\sc xxv} (we
ignore the weak features near 6.6-6.7 keV in $\gamma_{3}$ and
$\gamma_{5}$ because they do not show up in both of the HEG $\pm$1
orders). The absence of Fe\,{\sc xxv} is plausible from an ionization
perspective, given the high continuum flux in the $\gamma$ state. Our
multiplicative Gaussian model (Section \ref{sec:betahetgs}) indicates
that $N_{\rm Fe\,26}\sim6\times10^{17}$ cm $^{-2}.$ If all the iron is
locked up in Fe\,{\sc xxvi} (again, plausible because we do not
observe any other lines), this value of ${N_{\rm Fe\,26}}$ implies
an equivalent hydrogen column density of $N_{\rm H}\sim2\times10^{22}$ 
cm$^{-2}$. However, this hydrogen column should be regarded as a lower
limit, since there could be a significant population of fully-ionized
iron (which would increase the total iron column density
\textit{without} affecting the lines). 

As for the variability of the wind in the $\gamma$ state, it is clear
from Figure \ref{fig:gammaspec} that the continuum flux is variable,
but the lines themselves are difficult to distinguish because they
have similar significance, absorbed fluxes, and blueshifts. From Table
\ref{table:abs}, however, the Fe\,{\sc xxvi} line flux is highest at
intermediate luminosities, and the equivalent width drops at the
highest luminosities. We rule out a constant line flux at the
$\sim97\%$ confidence level. The variation of the equivalent width
between all three spectra is not particularly significant, but the
decrease in $W_{0}$ in $\gamma_{5}$ relative to $\gamma_{1}$ and
$\gamma_{3}$ has a formal significance of $\sim97\%.$ In general, the
lines do not appear to track the continuum flux directly, and so may
be subject to other influences (Section \ref{sec:discuss}).\vspace{-7mm} 

\section{DISCUSSION}
\label{sec:discuss}
In GRS 1915+105, the accretion disc wind is now known to vary on
time-scales ranging from a few seconds to (nearly) decades. This
variability appears particularly in the column density and ionization
state of the wind, but is also evident in the gas density, velocity,
and velocity dispersion. 
In the rest of this section, we discuss the
implications of our results for the complex relationship between the
properties of the wind and the luminosity of the central X-ray source,
which is responsible not only for driving the wind, but also for
ionizing it \citep[e.g.][]{B83,PK02,N11a}.

\subsection{The $\beta$ state: dip vs.\ flare}
\label{sec:dipflare}
First we consider the dip and flare intervals of the $\beta$ state
described in Section \ref{sec:betahetgs}. To recap briefly, we find
that when moving from the flaring phase into the hard X-ray dip, the
column density of the wind decreases significantly ($\sim3\times$),
while its velocity and velocity dispersion remain constant. 
Furthermore, the ionization state of the absorber does not appear to
change despite changes in the shape of the ionizing spectrum. We
argued in Section \ref{sec:betapca} that changes in the density or
geometry of the wind might therefore be important in determining the
properties of the absorber.

\begin{figure}
\centerline{\includegraphics[width=3.2 in]{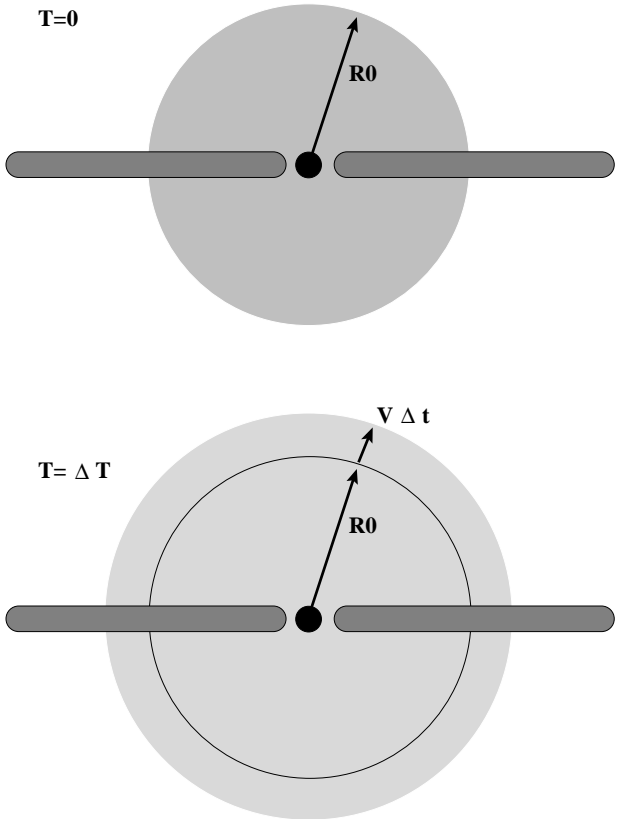}}
\caption{Toy model for the evolution of the column density in the
  $\beta$ state of GRS 1915+105. (TOP): the flaring interval, where
  the (spherical) wind has characteristic radius $R_{0}$ and a high
  column density (indicated by dark shading). (BOTTOM): during the
  dip, a time $\Delta T$ later, the wind has expanded and its column
  density has decreased accordingly (lighter shading). In Section
  \ref{sec:dipflare}, we use this model and our measurements of the
  column density to estimate $R_{0}\approx8\times10^{10}$ cm.}
\label{fig:geometry}
\end{figure}

For a toy model of this variability, suppose that the wind is launched 
from the disc as a sphere expanding with constant velocity ${\rm v},$
and suppose further that the wind is launched solely during the
bright, soft, X-ray flares. In other words, no wind is driven off the
disc during the hard X-ray dip. For simplicity, we assume the wind is 
spherical, but as long as its solid angle is roughly constant this toy
model will still apply. By the end of the flaring phase of the $\beta$
cycle, which we define as $T=0$, this sphere has a characteristic
radius $R_{0}$ that is comparable to the typical launch radius of the
wind (see the top panel of Figure \ref{fig:geometry}). During the
course of a flaring interval, the accretion disc launches $N$
particles into the wind, so that the expected column density is 
\begin{equation}
N_{\rm H0}=\frac{N}{4\pi R_{0}^{~2}}.
\end{equation}
During the dip, an average time $\Delta T$ later, when the
characteristic radius of the wind is
\begin{equation}
R_{1}=R_{0}+{\rm v}\Delta T,
\end{equation}
the expected column density is 
\begin{eqnarray}
N_{\rm H1}&=&\frac{N}{4\pi R_{1}^{~2}} \\
         &=&\frac{N}{4\pi (R_{0}+{\rm v}\Delta T)^{2}}.
\end{eqnarray}
Given the ratio of column densities in and out of the hard
X-ray dip, as well as the wind speed and $\Delta T$, we can solve for
the wind launch radius $R_{0}:$
\begin{equation}
\label{eqn:r0}
R_{0}=\frac{{\rm v}\Delta T}{\Phi-1},
\end{equation}where $\Phi=\sqrt{N_{\rm H0}/N_{\rm H1}}.$ Using the
values reported in Section \ref{sec:betahetgs}, Table \ref{table:abs},
and an average mid-flare to mid-dip $\Delta T=626$ s, we calculate
$R_{0}=(8\pm3)\times10^{10}$ cm. The error comes from propagating
$1\sigma$ uncertainties for each of the measured quantities through
Equation (\ref{eqn:r0}). The implied gas density in the wind is
$n_{e}=(1.4\pm0.5)\times10^{12}$ cm$^{-3},$ and the expected
ionization parameter is around $10^{5.2}$ ergs cm s$^{-1}.$
This is rather high for the observed lines, but could be reduced if
the wind is clumpy or accelerates in the dip, as hinted at in Table
\ref{table:abs}. The average mass loss rate in the typical $\sim800$ s
flare phase is $\dot{M}_{\rm wind}=(7\pm5)\times10^{18}$ g
s$^{-1}$; the instantaneous mass loss rate from the disc could be
significantly higher if the wind is only launched during a fraction of
the flare phase, or if the ionic abundance of completely-ionized iron
is not negligible during the dip. 

Of course, the fact that this toy model returns sensible results does
not imply or require that it is correct. We could alternately suppose,
for example, that the wind is produced at a constant rate throughout
the cycle. In this case, we can replace $\Phi$ in equation 5 with
$\chi=\Phi*\sqrt{1+\Delta T_{\rm dip}/\Delta T_{\rm flare}}.$ Here
the $\Delta T$s represent the duration of the dip and flare
intervals. For our data, $\Phi\sim1.7$ and $\chi\sim2.6$, so this
modification to our model reduces $R_{0}$ by a factor of $\sim2.$ The
density $n_{e}$ increases by a similar factor, but because the
ionization parameter $\xi$ scales with $R^{-2},$ the modified model
implies even more ionization than our original model (where the wind
shuts off in the dip). Overall, both are reasonable, but the excess
ionization in the modified (constant wind) model is potentially
problematic. Again, the ionization parameter could be reduced
if the wind is clumpy, but the modified model requires an even
clumpier wind than our original model, since the ionization parameters
are $\xi_{\rm modified}\approx2\xi_{\rm original}.$ Furthermore, we
know from previous work that hard X-ray spectra (similar to that
observed in the dip) tend to indicate activity in jets rather than
disc winds (e.g.\ \citealt{K02}; \citealt*{FBG04,NL09} and references
therein). This would lead us to the prediction that the hard dip would
be more likely to produce a jet than a wind. For these reasons, we
suggest that our original model for the wind variability in the
$\beta$ state is preferable, and that the wind actually does turn off
during the dip.   

As in our previous work on the persistent properties of the wind on
long time-scales and its variability on short time-scales
(\citealt{NL09,N11a}; see also \citealt{Luketic10}), our results are  
consistent with a wind driven primarily via a combination of Compton
heating of the outer accretion disc and radiation pressure on free
electrons. Furthermore, our results hint at a unifying process driving
the wind. In the heartbeat ($\rho$) state, which consists of bright
X-ray pulses every $\sim50$ seconds, we found that each X-ray pulse
heats up a large region of the disc and drives new gas into the line
of sight \citep{N11a}. Since the flares in the $\beta$ state are
similar to the pulses in the heartbeat state (as will be
demonstrated definitively in our forthcoming analysis of a new
\textit{Chandra,~RXTE,~EVLA} and \textit{Gemini} campaign; Neilsen et 
al.\ 2012, in preparation), the mechanisms producing the winds are
likely similar as well.

Yet the wind in the $\beta$ state also exhibits a critical difference
from the wind in the heartbeat state, for our toy model of the $\beta$
state suggests that the wind actually shuts off during the hard
X-ray dip. We find no evidence in the line ratios for over-ionization
of the absorber during the dip, and the ionizing flux is comparable
during both phases of the cycle. It seems improbable that such modest
changes in the ionizing and bolometric luminosities could lead to
enough over-ionization to reduce the apparent wind column density so
significantly. Thus we conclude that there is a long interval where
\textit{no wind is launched from the disc}, a fact that is rather
counterintuitive because the wind absorption lines never fade
completely. But during the dip, we actually detect the remnant of 
the wind launched by prior flares (in the form of
significantly weaker absorption lines).

This result poses an intriguing puzzle: why doesn't the disc
produce any wind during the X-ray dip? Our data do not provide any 
clear answers to this question, but the dip and the flares have
similar average X-ray luminosities, so it seems improbable that the
wind is quenched solely due to changes in the heating or energetics of
the outer disc. 

One possibility is that the highly-ionized wind in GRS 1915+105 is
primarily produced by short but intense pulses of radiation, as
observed in the heartbeat and $\beta$ states. If this impulsive
driving process launches partially- or temporarily-ballistic
parcels of gas into the line of sight, it might explain how the
Fe\,{\sc xxvi}/Fe\,{\sc xxv} ratio can remain roughly constant over a
large range in luminosity \citep[see][]{Kallman01}: ionization would only
become important as the wind rises well above the disc. Along this
line of reasoning, it is interesting that \citet{M06b} detected a wind
with very similar ionization properties in H1743--322 while the source
was undergoing moderate $\sim300$ s oscillations. Still, a general
link between wind formation, ionization, and X-ray variability leaves
much to be explained. For example, if variability is important, why
are these winds rarely observed during X-ray hard states, where the
rms X-ray variability is high? What determines whether the wind is
highly ionized, as in H1743--322 and the $\beta,~\gamma,$ and $\rho$
states of GRS 1915+105, or cooler with many lines, as in GX 13+1
\citep{U04}, GRO J1655-40 \citep{M06a}, and the $\phi$ state of GRS
1915+105 \citep{U09}? 

With these questions unanswered, a simpler explanation for the absence
of a wind is that the outer accretion disc is shadowed during the hard
X-ray dip and does not receive the full irradiating force of the X-ray
luminosity. This could be possible if, for example, the inner disc is
puffed up during the dip, so that low inclinations are shielded from
the X-ray emission (see also \citealt{Ponti11}). In
this explanation we find an exciting link to the formation of jets,
since it is believed that steady jets may require
geometrically-thick discs \citep[e.g.][]{Meier01}. To be clear, the
$\beta$ state produces discrete synchrotron blobs (baby jets;
\citealt{Fender97,PF97,M98}), not the compact, flat-spectrum jets for
which the thick disk theory was developed. However, a thick disk may
also be required to produce other kinds of jets, like those
observed after the hard dip in the $\beta$ state. Whether the
baby jet is launched continuously during the hard dip
\citep[e.g.][]{K02} or at the moment of the ``trigger spike''
\citep[e.g.][]{Rothstein05}, it seems reasonable to assume that the 
dip in the lightcurve coincides with changes in the structure of the
inner disc that ultimately lead to the ejection of a synchrotron
blob. These same changes may therefore hide a significant fraction of
the central X-ray luminosity from the wind-producing outer disc.

As for the relative timing of the disc and wind variability, we note
that the $\beta$ state is associated with periodic changes in the disc
mass accretion rate (e.g.\ \citealt{Migliari03}), whether or not their
origin is completely understood. It is expected that mass accretion
rate and scale height in the inner disc can evolve on viscous and
thermal time-scales, respectively (\citealt*{FKR02}). The
thermal time-scale is comparable to the dynamical time-scale, so the
scale height of the disc can change rapidly relative to the accretion
rate at any radius. To be quantitative, we consider the radius $R_{\rm
visc}$ at which the viscous time $t_{\rm visc}$ is equal to the
$\beta$ state's 30-minute cycle (i.e.\ we assume that the $\beta$
state is in part a viscous instability like many other states of GRS
1915+105; e.g.\ \citealt*{Nayakshin00}). Using Equation 1 of
\citet{B97b}, we find that $R_{\rm visc}\lesssim7\times10^{9}$ cm even
if $\dot{M}$ is as high as $10^{20}$ g s$^{-1}$ (assuming
a viscosity parameter $\alpha=0.01$ and black hole mass
$M=14~M_{\odot}$). At this radius, the thermal time-scale is $\sim0.1$
second. Thus any part of the disc participating in 30-minute viscous
variations can change its scale height almost instantly, and in any
case on time-scales much shorter than the duration of the X-ray dip.
Furthermore, we know that winds driven by thermal and radiation
pressure can respond to changes in the luminosity in as few as five
seconds \citepalias{N11a}. Therefore, we conclude that our
shadowing explanation is plausible.

If this  scenario is correct, it would explain the uncanny
ability of the wind to ``know'' about the state of the inner disc,
i.e.\ why winds might be preferentially formed when jets are
absent. Because the radiation from the disc launches the wind, the
wind can effectively respond to changes in the driving luminosity
\textit{as they occur}. There may be delays due to light travel time,
but this is only a few seconds and is short compared to the duration
of the dip. In short, the remarkable quenching of the wind during the 
hard X-ray dip is a clear consequence of the radiative links between
the inner and outer parts of the accretion flow, consistent with our
results for the heartbeat state on time-scales of 5 seconds
\citep{N11a}.\vspace{-4mm}
\begin{figure}
\centerline{\includegraphics[width=3.2 in]{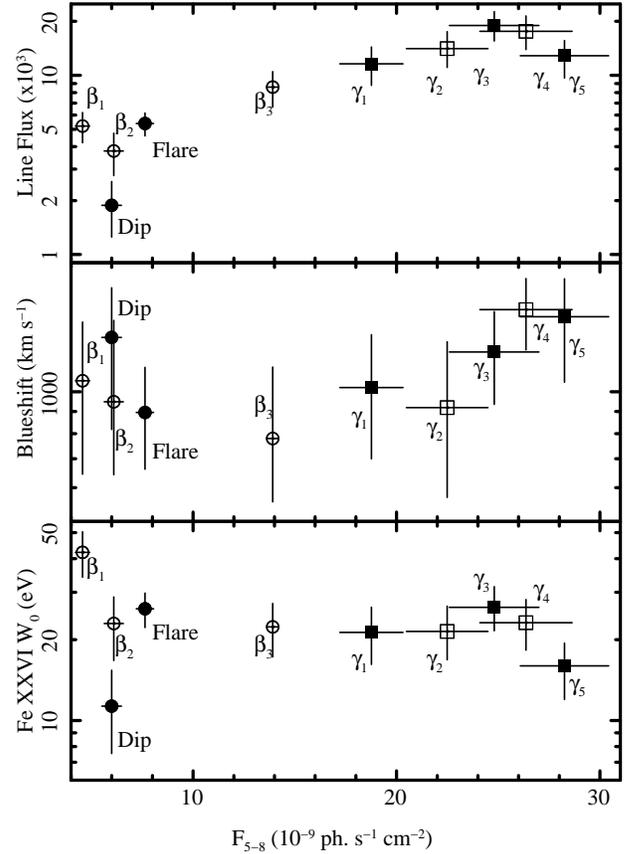}}
\caption{Flux-resolved properties of the accretion disc wind in the
  $\beta$ and $\gamma$ states of GRS 1915+105. The $\beta$ state
  points are marked with circles, while the results for the $\gamma$
  state are marked with squares. For a given state, symbols with the
  same fill style are statistically independent. Top panel: the
  Fe\,{\sc xxvi} line flux, which generally increases with increasing
  continuum flux. Middle panel: the wind blueshift, which is
  consistent with a constant over all flux. Bottom panel: the 
  Fe\,{\sc xxvi} equivalent width, which is mostly constant; see
  Section \ref{sec:discuss} for a discussion of deviations from a
  constant equivalent width.}
\label{fig:linesvsflux}
\end{figure}
\subsection{Flux-resolved studies}
The complex radiative links between the wind and the inner and outer
accretion disc are also apparent in the flux dependence of the
properties of the wind absorption lines. For the $\beta$ and $\gamma$
states, we summarize our results in Figure \ref{fig:linesvsflux}. Note
that any two points in Figure \ref{fig:linesvsflux} that have the same
symbol and fill style are statistically independent. At this level, we
do not have the signal to statistically distinguish changes in the
wind speed, but this figure clearly demonstrates that the line flux
and equivalent width exhibit a dependence on the observed continuum
flux that is non-linear and not necessarily even monotonic. 

Part of this non-linearity is due to the behavior discussed in Section
\ref{sec:dipflare}, in which residual absorption lines from a wind
launched in some previous epoch may still be present in the X-ray
spectrum. As discussed in the preceding section, wind driving may be
sensitive to both the X-ray luminosity and the state and structure of
the inner accretion flow.

For the most part, the line flux rises smoothly and
the equivalent width is remarkably constant over a factor of nearly
five in continuum flux. This would seem to indicate that the 
Fe\,{\sc xxvi} column density changes very slowly with luminosity,
even across different states. A direct comparison between
$\beta$ and $\gamma$ is difficult because of possible changes in
line saturation, and because the lines are generally resolved in the  
$\gamma$ state but generally not in the $\beta$ state. Nevertheless,
the smooth evolution of the wind properties suggests a common origin
(i.e.\ radiative/thermal driving), in which case we can expect the 
accretion disc wind and the continuum flux to be strongly linked.  

For example, the weakness or absence of Fe\,{\sc xxv} in the $\gamma$
spectra (relative to $\beta$) does suggest that as the X-ray
luminosity increases, the wind does become progressively more
ionized. This conclusion is supported by the behavior of the 
Fe\,{\sc xxv}/Fe\,{\sc xxvi} ratio in our flux-resolved $\beta$ flare
spectra. In addition, because the abundance of Fe\,{\sc xxvi} should
drop sharply due to over-ionization above some critical luminosity, we
believe the drop in the Fe\,{\sc xxvi} line flux and equivalent width
after $\gamma_{3}$ (Figure \ref{fig:linesvsflux}) further reinforce
the conclusion that the X-ray luminosity is a major determining factor
of the wind properties. 

On the other hand, it remains clear that there is more to disc winds
than the luminosity that drives them. For example, the density of the
wind in GRS 1915+105 seems to be independent of luminosity at around
$10^{12}$ cm$^{-3}$ (\citealt{L02,U09,N11a}; this work). Or consider
that the equivalent width of the iron line is highest in $\beta_{1},$
which has the lowest continuum flux studied in this work. It may be
that different driving mechanisms are important at different
luminosities, i.e.\ MHD winds \citep{Proga2000,M06a,M08}, or it may be
that flux-resolved studies like the one presented here fail to capture
the true time-dependence of variable winds. In the heartbeat state,
the wind is launched \textit{by} high luminosities but doesn't rise
into our line of sight until after the X-ray luminosity has fallen 
\citepalias{N11a}; the same may also be true during the $\beta$
flares and the $\gamma$ state, whose variability is even more
erratic.\vspace{-7mm} 

\section{CONCLUSIONS}
\label{sec:conc}
In this paper, we have presented a time- and flux-resolved
high-resolution spectral analysis of GRS 1915+105 in the $\beta$ and
$\gamma$ states. Although only the $\beta$ state is associated with
transient jet formation, both states exhibit strong erratic X-ray
variability, which we have used to probe the relationship between the
central X-ray luminosity and the properties of the accretion disc
wind. In particular, our new results confirm our previous finding that 
radiation mediates rapid interactions between the inner accretion 
disc and the outer accretion disc. This interaction is most evident
in the $\beta$ state, where the wind is actually quenched during the
hard X-ray dip, possibly because the irradiation of the outer disc is
affected by changes in the structure of the inner disc leading to the
formation of a transient jet. To the best of our knowledge, this is
the first time that a wind has been observed to turn off within a
single observation, and the mechanism ($\dot{M}$ and/or structural
changes in the inner disc) requires intricate links between radiation,
winds, and jets.

This result is therefore an exciting step towards a complete
understanding of wind-jet interactions on all time-scales. With
our sensitivity and limited time resolution, our best estimates place
the wind in the $\beta$ state $\sim10^{11}$ cm from the black hole,
leaving it effectively unable to affect the transient jet on
such short time-scales. However, if future multiwavelength studies
unaffected by telemetry saturation are able to place better
constraints on the location of the wind, and require it to be very
close to the black hole, then the $\beta$ state might produce the
first evidence for direct wind-jet interactions on any time
scale. Such a result would significantly enhance our understanding of
accretion and ejection as profoundly inter-related processes around
black holes.\vspace{-7mm} 

\section*{acknowledgements} We thank the referee for comments that
improved the clarity of our presentation. We thank Ron Remillard and
Claude Canizares for helpful discussions of spectral
variability. J.N.\ and J.C.L.\ gratefully acknowledge funding support
from \textit{Chandra} grant AR0-11004X, the Harvard University
Graduate School of Arts and Sciences, and the Harvard University
Faculty of Arts and Sciences. J.N.\ acknowledges additional support
from the National Aeronautics and Space Adminisration through the
Smithsonian Astrophysical Observatory contract SV3-73016 to MIT for
support of the \textit{Chandra} X-ray Center, which is operated by the
Smithsonian Astrophysical Observatory for and on behalf of the
National Aeronautics Space Administration under contract
NAS8-03060.\vspace{-7mm}  

\bibliographystyle{mn}
\bibliography{mn}

\bsp

\label{lastpage}

\end{document}